\documentclass[twocolumn,10pt]{IEEEtran}
\usepackage{ifpdf, flushend}

\ifCLASSINFOpdf
  \usepackage[pdftex]{graphicx}
  \graphicspath{{../pdf/}{../jpeg/}}
  \DeclareGraphicsExtensions{.pdf,.jpeg,.png}
\else
  \usepackage[dvips]{graphicx}
  \graphicspath{{../eps/}}
  \DeclareGraphicsExtensions{.eps}
\fi
\usepackage[cmex10]{amsmath}
\usepackage {amssymb}
\usepackage{algorithmic}
\usepackage{array}
\usepackage{mdwmath}
\usepackage{mdwtab}
\usepackage{eqparbox}
\usepackage{url}
\usepackage{hyperref}
\usepackage{algorithm}
\usepackage{algorithmic}

\newcommand{\argmax}{\operatornamewithlimits{argmax}}
\newcommand{\beq}{\begin{equation}}
\newcommand{\eeq}{\end{equation}}
\newcommand{\beqn}{\begin{eqnarray}}
\newcommand{\eeqn}{\end{eqnarray}}
\newcommand{\beqno}{\begin{eqnarray*}}
\newcommand{\eeqno}{\end{eqnarray*}}
\newcommand{\bma}{\begin{displaymath}}
\newcommand{\ema}{\end{displaymath}}
\newcommand{\bnu}{\begin{enumerate}}
\newcommand{\enu}{\end{enumerate}}
\newcommand{\bce}{\begin{center}}
\newcommand{\ece}{\end{center}}
\newcommand{\btb}{\begin{tabular}}
\newcommand{\etb}{\end{tabular}}

\begin{document}

%
\title{Distributed MAC Protocol Design for Full--Duplex Cognitive Radio Networks}

\author{\IEEEauthorblockN{Le Thanh Tan and Long Bao Le}  
\thanks{The authors are with INRS-EMT, University of Quebec,  Montr\'{e}al, Qu\'{e}bec, Canada. 
Emails: \{lethanh,long.le\}@emt.inrs.ca.}}

\maketitle

\begin{abstract}
\boldmath
In this paper, we consider the Medium Access Control (MAC) protocol design for full-duplex cognitive radio networks (FDCRNs). Our design exploits the fact that 
full-duplex (FD) secondary users (SUs) can perform spectrum sensing and access simultaneously, which enable them to detect the primary users' (PUs) activity during transmission. 
The developed FD MAC protocol employs the standard backoff mechanism as in the 802.11 MAC protocol. However, we propose to adopt the frame fragmentation
during the data transmission phase for timely detection of active PUs where each data packet is divided into multiple fragments and the active SU 
makes sensing detection at the end of each data fragment. Then, we develop a mathematical model to analyze the throughput performance of the proposed FD MAC protocol. 
Furthermore, we propose an algorithm to configure the MAC protocol so that efficient self-interference management and sensing overhead control can be achieved.
Finally, numerical results are presented to evaluate the performance of our design and demonstrate the throughput enhancement compared to
the existing half-duplex (HD) cognitive MAC protocol.
\end{abstract}

\begin{IEEEkeywords}
MAC protocol, spectrum sensing, optimal sensing, throughput maximization, full-duplex cognitive radios.
\end{IEEEkeywords}
\IEEEpeerreviewmaketitle

\section{Introduction}

Engineering MAC protocols for efficient sharing of white spaces is an important research problem for cognitive radio networks (CRNs).
In general, a cognitive MAC protocol must realize both the spectrum sensing and access functions so that timely detection of the PUs' activity 
and effective spectrum sharing among different SUs can be achieved. Most existing research works on cognitive MAC protocols have focused 
on the design and analysis of half-duplex  CRNs (e.g., see \cite{Yu09, Cor09} and references therein).
Due to the half-duplex constraint, SUs typically employ the two-stage sensing/access procedure where they sense the spectrum in the first stage before accessing
available channels for data transmission in the second stage \cite{Liang08} -- \cite{Konda08}.   
These HD MAC protocols may not exploit the white spaces very efficiently since significant sensing time can be required, which
would otherwise be utilized for data transmission. Moreover, SUs may not timely detect the PUs' activity during data transmission, which causes 
 severe interference to active PUs.

Thanks to recent advances in the full-duplex technologies, some recent works propose more efficient full-duplex (FD) spectrum access design for cognitive radio networks
 \cite{ Afifi14} where each SU can perform sensing and transmission simultaneously \cite{Duarte12}. 
In general, the self-interference\footnote{Self-interference is due to the power leakage from the transmitter
to the receiver of a full-duplex transceiver.} due to simultaneous sensing and access may lead to degradation on the SUs'
spectrum sensing performance. 
In \cite{Afifi14}, the authors consider the cognitive FD MAC design where they assume that SUs perform sensing in multiple small time slots
to detect the PU's activity during transmission, which may not be efficient. Moreover, they assume that the PU can change its idle/busy
status at most once during the SU's transmission, which may not hold true if the SU's data packets are long. 
Our FD cognitive MAC design overcomes these limitations where we propose
to employ frame fragmentation with appropriate sensing design for timely protection of the PU and
we optimize the sensing duration to maximize the network throughput. 

Specifically, our FD MAC design employs the standard backoff mechanism as in the 802.11 MAC protocol to solve contention among SUs for compatibility.
However, the winning SU of the contention process performs simultaneous sensing and transmission during the access phase
where each data packet is divided into multiple data fragments and sensing decisions are taken at the end of individual
data fragments. This packet fragmentation enables timely detection of PUs since the data fragment time is chosen to be smaller than the required
channel evacuation time. We then develop a mathematical model for throughput performance analysis of the proposed FD MAC design
considering the imperfect sensing effect. Moreover, we propose an algorithm to configure different design parameters
including data fragment time, SU's transmit power, and the contention window to achieve the maximum throughput.
Finally, we present numerical results to illustrate the impacts of different protocol parameters on the throughput performance
and the throughput enhancement compared to the existing HD MAC protocol.

 
\section{System and PU Activity Models}
\label{SystemModel}


\subsection{System Model}
\label{System}

We consider a network setting where $n_0$ pairs of SUs opportunistically exploit white spaces on a frequency band for data transmission. 
We assume that each SU is equipped with one full-duplex transceiver, which can perform sensing and transmission simultaneously.
However,  SUs suffer from self-interference from its transmission during sensing (i.e., transmitted signals are leaked into
the received PU signal). We denote $I$ as the average self-interference power where it can be modeled as
 $I = \zeta \left(P_s\right)^{\xi}$ \cite{Duarte12} where $P_s$ is the SU transmit power, $\zeta$ and $\xi$ ($0 \leq \xi \leq 1$) are some
predetermined coefficients which capture the self-interference cancellation quality.
We design a asynchronous MAC protocol where no synchronization is assumed between SUs and PUs.
We assume that different pairs of SUs can overhear transmissions from the others (i.e., collocated networks). 
In the following, we refer to pair $i$ of SUs as secondary link $i$ or flow $i$ interchangeably.

\subsection{Primary User Activity}
\label{PUAM}

We assume that the PU's idle/busy status follows two independent and identical distribution processes. Here the channel is available and busy 
for the secondary access if the PU is in the idle and busy states, respectively. Let $\mathcal{H}_0$ and $\mathcal{H}_1$ denote the events 
that the PU is idle and active, respectively. To protect the PU, we assume that SUs must stop their transmission and evacuate from the channel 
within the maximum delay of $T_{\sf eva}$, which is referred to as channel evacuation time. 

Let $\tau_{\sf ac}$ and $\tau_{\sf id}$ denote the random variables which represent the durations of channel active and idle states, respectively.
We assume that $\tau_{\sf ac}$ and $\tau_{\sf id}$ are larger than $T_{\sf eva}$.
We denote probability density functions of $\tau_{\sf ac}$ and $\tau_{\sf id}$ as $f_{\tau_{\sf ac}}\left(t\right)$ and $f_{\tau_{\sf id}}\left(t\right)$, respectively.  
In addition, let  $\mathcal{P}\left(\mathcal{H}_0\right) = \frac{{\bar \tau}_{\sf id}}{{\bar \tau}_{\sf id}+{\bar \tau}_{\sf ac}}$ and  $\mathcal{P}\left( \mathcal{H}_1 \right) = 
1 - \mathcal{P}\left(\mathcal{H}_0\right)$ present the probabilities that the channel is available and busy, respectively.

\section{MAC Protocol Design}
\label{SingleChan}

\begin{figure}[!t]
\centering
\includegraphics[width=75mm]{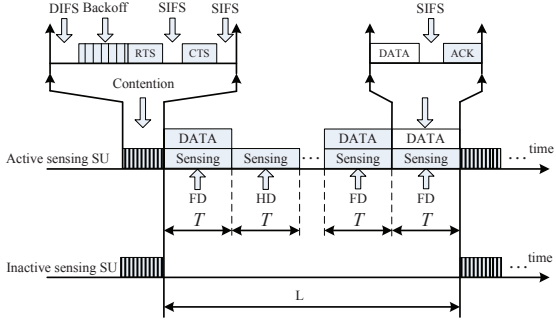}
\caption{Timing diagram of the proposed full-duplex MAC protocol.}
\label{Sentime}
\end{figure}

For contention resolution, we assume that SUs employ the backoff as in the standard CSMA/CA protocol \cite{Bian00}.
In particular, SU transmitters perform carrier sensing and start the backoff after the channel is sensed to be idle in an interval referred to as
 DIFS (DCF Interframe Space). Specifically, each SU chooses a random backoff time, which is uniformly distributed in the range $[0,2^i W-1]$, $0 \leq i \leq m$
and starts counting down while carrier sensing the channel where $W$ denotes the minimum backoff window and $m$ is the maximum backoff stage.
For simplicity, we assume that $m$=1 in the throughput analysis in the next section and  we refer to $W$ simply as the contention window.
 
Let $\sigma$ denote a mini-slot interval, each of which corresponds one unit of the backoff time counter. Upon hearing a transmission from other secondary links or PUs, all SUs will ``freeze''
its backoff time counter and reactivate when the channel is sensed idle again. Otherwise, if the backoff time counter reaches zero, the underlying secondary link wins the contention. 
To complete the reservation, the four-way handshake with Request-to-Send/Clear-to-Send (RTS/CST) exchanges will be employed to reserve the available channel for transmission in the next stage.
Specifically, the secondary transmitter sends RTS to the secondary receiver and waits until it successfully receives CTS from the secondary receiver before sending the data. 
All other SUs, which hear the RTS and CTS from the secondary winner, defer to access the channel for a duration equal to the data packet length, $L$.
Furthermore, the standard small interval, namely SIFS (Short Interframe Space), is used before the transmissions of CTS, ACK and data frame as described in \cite{Bian00}.

We assume that the data length of the SU transmitter, $L$ ($L = \overline{K} T$) is larger than the evacuation time, $T_{\sf eva}$. Hence, SUs divide each packet 
into $\overline{K}$ equal-size data fragments, each of which is transmitted in duration $T$. Moreover,  $T$ is chosen to be smaller than $T_{\sf eva}$ so
that timely evacuation from the busy channel can be realized.
In addition, the active SU transmitter simultaneously senses the PU activity and transmits its data in each fragment
where it makes one sensing decision on the idle/active channel status at the end of each data fragment.
Furthermore, if the sensing outcome at one particular fragment indicates an ``available'' channel then
the active SU transmitter performs concurrent sensing and transmission in the next fragment
(called full-duplex (FD) sensing); otherwise, it only performs sensing without transmission (referred to as half-duplex (HD) sensing). 
This design allows to protect the PU with evacuation delay at most $T$ if the sensing is perfect. 
We assume that SU's transmit power is set equal to $P_s$ where we will optimize this parameter to achieve
good tradeoff between  self-interference mitigation and high communication rate later. 
The timing diagram of this proposed FD MAC protocol is illustrated in Fig.~\ref{Sentime}.

\section{Throughput Analysis}
\label{Tput_Ana}

We perform throughput analysis for the saturated system where all SUs are assumed to always have data to transmit \cite{Bian00}. 
Let $\mathcal{P}_{\sf succ} \left(i_0\right)$ denote the probability that the SU successfully reserves the channel (i.e., the RTS and CTS are exchanged successfully), $T_{\sf ove} \left(i_0\right)$ represent  the time overhead due to backoff and RTS/CTS exchanges, and $\mathcal{T}^{i_0}$ denote the average conditional throughput
in bits/Hz for the case where the backoff counter of the winning SU is equal to  $i_0$ ($i_0 \in \left[0, W-1\right]$). Then,
 the normalized throughput can be written as 
\beqn
\label{EQN_T_put}
\mathcal{NT} = \sum_{i_0 = 0}^{W - 1} \mathcal{P}_{\sf succ} \left(i_0\right) \times \frac{\mathcal{T}^{i_0}}{T_{\sf ove} \left(i_0\right) + \overline{K}T}.
\eeqn
In this expression, we have considered all possible values of the backoff counter of the winning SU in $\left[0, W-1\right]$.
In what follows, we derive the quantities $\mathcal{P}_{\sf succ} \left(i_0\right)$, $T_{\sf ove} \left(i_0\right)$, and $\mathcal{T}^{i_0}$.

\subsection{Derivations of $\mathcal{P}_{\sf succ} \left(i_0\right)$ and $T_{\sf ove} \left(i_0\right)$}


The event that the SU successfully reserves the channel for the case the winning SU has its backoff counter equal to $i_0$ occurs if
all other SUs choose their  backoff counters larger than $i_0$.
So the probability of this event ($\mathcal{P}_{\sf succ} \left(i_0\right)$) can be expressed as follows:
\beqn
\mathcal{P}_{\sf succ} \left(i_0\right) = n_0 \frac{1}{W} \left(\frac{W-1-i_0}{W}\right)^{n_0-1}.
\eeqn
Moreover, the corresponding overhead involved for successful channel reservation at backoff slot $i_0$ is 
\beqn
T_{\sf ove} \left(i_0\right) = i_0 \times \sigma + 2 SIFS + RTS + CTS + DIFS
\eeqn
where $\sigma$, $SIFS$, $DIFS$, $RTS$ and $CTS$ represent the duration of backoff slot, the durations of one SIFS, one DIFT, $RTS$, and $CTS$ control packets, respectively.

\subsection{Derivation of $\mathcal{T}^{i_0}$}

The quantity $\mathcal{T}^{i_0}$ can be derived by studying the transmission phase which spans $\overline{K}$ data fragment intervals each with length $T$.
Note that the PU's activity is not synchronized with the SU's transmission; therefore, the PU can change its active/idle status any time. 
It can be verified that there are four possible events related to the status changes of the PU during any particular data fragment, which are defined as follows.   
Let $\mathcal{H}_{00}$ be the event that the PU is idle for the whole fragment interval; $\mathcal{H}_{10}$ denote the event that PU is first active and then becomes idle by the
end of the fragment; $\mathcal{H}_{11}$ be the event that the PU is active for the whole fragment; and finally, $\mathcal{H}_{01}$ 
capture the event that PU is first idle and then becomes active by the end of the fragment. 
Here, there can be at most one transition between the active and idle states during one fragment time. 
This holds because we have $\tau_{\sf ac}$ and $\tau_{\sf id}$ are larger than the fragment time $T$ (since $T_{\sf eva}$ is larger than $T$;  
$\tau_{\sf ac}$ and $\tau_{\sf id}$ are larger than $T_{\sf eva}$).

The average throughput achieved by the secondary network depends on the PU's activity and sensing outcomes at every fragment.
For any particular fragment, if the sensing outcome indicates an available channel then the winning SU will perform concurrent transmission and 
sensing in the next fragment (FD sensing); otherwise, it will perform sensing only in the next fragment (HD sensing) and hence the achievable throughput is zero.
Moreover, the throughput and sensing outcome also depend on how the PU changes its state at one particular fragment.

In what follows, we present the steps to calculate $\mathcal{T}^{i_0}$.
We first generate all possible patterns capturing how the PU changes its idle/active status over all $\overline{K}$ fragments. 
For each generated pattern, we then consider all possible sensing outcomes in all fragments. 
Moreover, we quantify the achieved throughput conditioned on individual cases with corresponding PU's statuses and sensing 
outcomes in all $\overline{K}$ fragments based on which the overall average throughput can be calculated.

\begin{figure}[!t]
\centering
\includegraphics[width=75mm]{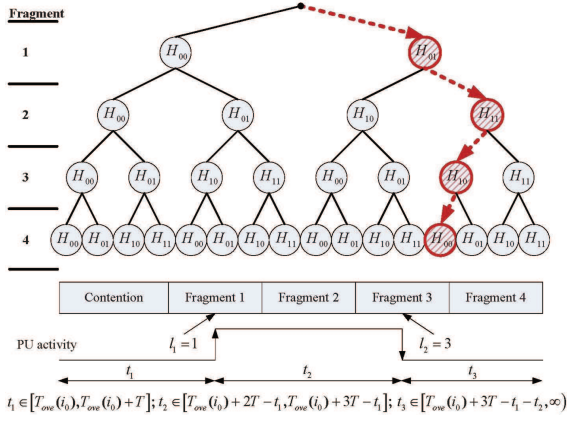}
\caption{PU's activity patterns.}
\label{List_scenario}
\end{figure}

Let $\mathcal{S}_i$ denote one particular pattern $i$ capturing the corresponding changes of the PU's idle/active status
and $\mathcal{AS}$ denote the set of all possible patterns. There are $2^{\overline{K}}$ possible patterns since the PU
either changes or maintains the status (idle or active status) in each fragment. Note that each pattern  $\mathcal{S}_i$
comprises a sequence of state changes represented by possible events $\mathcal{H}_{ij}$. For convenience, we define
a pattern $\mathcal{S}_i$ by the corresponding four sets $\mathcal{S}_i^{00}$, $\mathcal{S}_i^{10}$, $\mathcal{S}_i^{01}$, and $\mathcal{S}_i^{11}$ 
whose elements are fragment indexes during which the channel state changes corresponding to $\mathcal{H}_{00}$, $\mathcal{H}_{10}$, $\mathcal{H}_{01}$ and $\mathcal{H}_{11}$ occur,
respectively. For example, if one pattern has the
PU's state changes as $\mathcal{H}_{00}$, $\mathcal{H}_{01}$, $\mathcal{H}_{11}$, $\mathcal{H}_{11}$  in this order
then we have $\mathcal{S}_i^{00}=\left\{1\right\}$,  $\mathcal{S}_i^{10} = \emptyset$, $\mathcal{S}_i^{01}=\left\{2\right\}$, and $\mathcal{S}_i^{11}=\left\{3, 4 \right\}$.
Fig.~\ref{List_scenario} shows all possibles patterns for $\overline{K} = 4$.
Then, the conditional throughput $\mathcal{T}^{i_0}$ can be written as
\beqn
\label{EQN_T_T}
\mathcal{T}^{i_0} = \sum _{i=1}^{\left|\mathcal{AS}\right|} \mathcal{T}^{i_0} \left\{\mathcal{S}_i\right\} 
\eeqn
where $\mathcal{T}^{i_0} \left\{\mathcal{S}_i\right\}$ is the throughput (bits/Hz) for the pattern $\mathcal{S}_i$. 


Consider one particular pattern $\mathcal{S}_i$. Let $\Omega_i$ be the set with cardinality $\rho_i$ whose elements are fragment indexes $l_j^i$ in which the PU changes its state.
Moreover, let $\Gamma_i$ be another set also with cardinality $\rho_i$ whose elements $t_j^i$ are time intervals between consecutive PU's state changes. 
In the following, we omit index $i$ in parameters $l_j^i$ and $t_j^i$ for brevity.
We show one particular pattern $\mathcal{S}_i$ with the parameter $t_j$ in Fig.~\ref{List_scenario} (the corresponding $\mathcal{S}_i$ is indicated by dash lines).
For convenience, we denote $\int_{t=a}^b$ as $\int_{t\in \mathcal{R}}$ where $\mathcal{R} \equiv \left[a,b\right]$ is the range of $t$.
Moreover, we define $\mathcal{R}_1 = \left[\! T_{\sf ove} \left(i_0\right)\!+\!(l_1\!-\!1)T, T_{\sf ove} \left(i_0\right)+l_1T\right]$ to be the range for $t_1$. 
Similarly, the range for $t_j$ is $\mathcal{R}_j = \left[T_{\sf ove} \!\left(i_0\right)\!+\!(l_j\!-\!1)T\!-\!\sum_{r=1}^{j-1}t_r, T_{\sf ove} \left(i_0\right)\!+\!l_jT-\sum_{r=1}^{j-1}t_r\right]$;
then the range for $t_{\rho_i}$ can be also expressed as $\mathcal{R}_{\rho_i} = \left[T_{\sf ove} \!\left(i_0\right)\!+\!(l_{\rho_i}\!-\!1)T\!-\!\sum_{r=1}^{\rho_i\!-\!1}\!t_r, T_{\sf ove} \!\left(i_0\right)\!+\!l_{\rho_i}T\!-\!\sum_{r=1}^{\rho_i\!-\!1}\!t_r\right]$;
and finally the range for $t_{\rho_i+1}$ is written as $\mathcal{R}_{\rho_i+1} = \left[T_{\sf ove} \left(i_0\right)+\overline{K}T-\sum_{r=1}^{\rho_i}t_r, \infty\right)$.

Using these notations, $\mathcal{T}^{i_0} \left\{\mathcal{S}_i\right\}$ can be written as 
\beqn 
\mathcal{T}^{i_0} \left\{\mathcal{S}_i\right\} = \mathcal{P}\left(\mathcal{H}_0\right) \int\limits_{t_1 \in \mathcal{R}_1} \!\!\ldots\!\! \int\limits_{t_j \in \mathcal{R}_j} \!\!\ldots \!\!\int\limits_{t_{\rho_i} \in \mathcal{R}_{\rho_i}} \int\limits_{t_{\rho_i+1} \in \mathcal{R}_{\rho_i+1}} \nonumber\\
\sum_{k=1}^{2^{\overline K}} \prod_{j_1 \in \Phi_k^0} \mathcal{P}_{j_1} (\vec{t}^i) \prod_{j_2 \in \Phi_k^1} \mathcal{\overline P}_{j_2} (\vec{t}^i) 
\sum_{j_3 \in {\overline \Phi}_k^0} T_{j_3} (\vec{t}^i) \hspace{1.5cm}  \label{sensout} \\
f_{t_1}\!(\!t_1\!) \!\!\ldots \!\!f_{t_j}\!(\!t_j\!) \!\!\ldots\!\! f_{t_{\rho_i}}\!(\!t_{\rho_i}\!) f_{t_{\rho_i+1}}\!(\!t_{\rho_i+1}\!) dt_1 \!\!\ldots\!\! dt_j\!\ldots\! dt_{\rho_i} dt_{\rho_i+1} \label {T_set_1}
\eeqn
where $\bar{x} = 1 - x$ and $f_{t_j}(t_j)$ in (\ref{T_set_1}) is the pdf of $t_j$, which can be either $f_{\tau_{\sf id}}(.)$ or $f_{\tau_{\sf ac}}(.)$ depending
on whether the underlying interval is idle or active one, respectively.

In this expression, we have averaged over the possible distribution of the time interval vector $\vec{t}^i$ whose elements ${t}^i_j$ vary according to
the exact state transition instant within the corresponding data fragment. The quantity in (\ref{sensout}) accounts for $2^{\overline K}$ different possible
sensing outcomes in ${\overline K}$ fragments. Moreover, the term $\sum_{j_3 \in {\overline \Phi}_k^0} T_{j_3} (\vec{t}^i)$ represents the corresponding
achieved throughput for the underlying pattern $\mathcal{S}_i$ and sensing outcomes. 
Note that the data phase must start with the idle state, which explains the factor
$\mathcal{P}\left(\mathcal{H}_0\right)$ in this expression. Moreover, the first channel transition must be from idle to active, i.e., $f_{t_1}(t_1) = f_{\tau_{\sf id}}(t_1)$. In general, if $j$ is odd, then $f_{t_j}(t_j) = f_{\tau_{\sf id}}(t_j)$; otherwise, $f_{t_j}(t_j) = f_{\tau_{\sf ac}}(t_j)$. 

We now interpret the term in (\ref{sensout}) in details. For particular sensing outcomes in all fragments, we have defined two subsets, namely a set of fragments $\Phi_k^0$ where the sensing 
 indicates an available channel and the complement set of fragments $\Phi_k^1$ where the sensing indicates a busy channel. Moreover, we have defined the set $\overline{\Phi}_k^0$ whose
elements are indices of fragments each of which is the next fragment of one corresponding fragment in $\Phi_k^0$ (e.g., if $\Phi_k^0 = \left\{1,3\right\}$ then 
$\overline{\Phi}_k^0 = \left\{2,4\right\}$). We need the set $\overline{\Phi}_k^0$ since only fragments in this set involve data transmissions and, therefore, contribute
to the overall network throughput. In (\ref{sensout}), the first two products represent the probability of sensing outcomes for all fragments in the $k$-th sensing outcome.


In the following, we present the derivations of $\mathcal{P}_j$ and $T_j$. 
Let  $\mathcal{P}_d^{11}$, $\mathcal{P}_d^{01}$, $\mathcal{P}_f^{00}$, and $\mathcal{P}_f^{10}$ denote the probabilities of detection and false alarm for
the following channel state transition events $\mathcal{H}_{11}$, $\mathcal{H}_{01}$, $\mathcal{H}_{00}$ and $\mathcal{H}_{10}$ using FD sensing, respectively.
Similarly, we denote $\mathcal{P}_{d,h}^{11}$ and $\mathcal{P}_{d,h}^{01}$, $\mathcal{P}_{f,h}^{00}$ and $\mathcal{P}_{f,h}^{10}$ as the probabilities of detection and false alarm for events $\mathcal{H}_{11}$, $\mathcal{H}_{01}$, $\mathcal{H}_{00}$ and $\mathcal{H}_{10}$ using HD sensing, respectively. 
Derivations of the probabilities of detection and false alarm for all events and all sensing schemes are given in Appendix \ref{CAL_P_F_P_D}.
Let  $T^{00}$, $T^{10}$, $T^{01}$ and $T^{11}$ denote the number of bits per Hz under the state transition events $\mathcal{H}_{00}$, $\mathcal{H}_{10}$, $\mathcal{H}_{01}$ and $\mathcal{H}_{11}$, respectively. These quantities are derived in Appendix \ref{TPUT_FRAGMEN}.

For particular fragment $j \in \Phi_k^0$, the quantities $\mathcal{P}_j$ and $T_j$ are given in Table \ref{table} where $\bigcap$ and $\bigcup$ denote  AND and OR operations, respectively.
For example, in the first line, if the sensing outcome indicates an available channel in fragment $j-1$ ($\left(j-1\right) \in \Phi_k^0$) or fragment $j-1$ is the first one
in the data access phase then the SU will perform concurrent sensing and transmission in fragment $j$. Moreover, fragment $j$ belongs to the set
$\mathcal{S}^{00}$; therefore, we have $\mathcal{P}_j = \mathcal{\overline P}_f^{00}$ and $T_j = T^{00}$. Similarly, we can interpret the results for
 $\mathcal{P}_j$ and $T_j$ in other cases. 
For fragment $j \in \Phi_k^1$, to calculate the quantities $\mathcal{ P}_j$, we use the results in the first column of Table ~\ref{table} but
we have to change all items to $\mathcal{ P}_x^{kl}$ ($k,l \in \left\{0,1 \right\}$, $x$ represents $f$, $d$, $fh$, and $dh$).
Note that all the quantities depends on time instant $t$ when the PU changes its state. For example, at the time point $t_j$ corresponding to fragment $l_j$, $t = \sum_{k=1}^j t_k - \left[T_{\sf ove} \left(i_0\right)+(l_j-1)T\right]$.
We have omitted this dependence on $t$ in all notations for brevity (details can be found in Appendices ~\ref{CAL_P_F_P_D} and \ref{TPUT_FRAGMEN}).

\begin{table}
\tiny
\centering
\caption{Determination of $\mathcal{P}_j$ and $T_j$ for fragment $j \in \Phi_k^0$}
\label{table}
\begin{tabular}{|c|c|c|}
\hline 
 $\mathcal{P}_j$ & $T_j$ & Conditions \tabularnewline
\hline
\hline 
 $\mathcal{\overline P}_f^{00}$ & $T^{00}$ & $\left(j\in \mathcal{S}^{00}\right) \bigcap \left\{\left[\left(j-1\right) \in \Phi_k^0\right]\bigcup\left(j-1=1\right)\right\}$ \tabularnewline
\hline 
 $\mathcal{\overline P}_{f,h}^{00}$ & 0 & $\left(j\in \mathcal{S}^{00}\right) \bigcap \left[\left(j-1\right) \in \Phi_k^1\right]$  \tabularnewline
\hline 
 $\mathcal{\overline P}_f^{10}$ & $T^{10}$  & $\left(j\in \mathcal{S}^{10}\right) \bigcap \left\{\left[\left(j-1\right) \in \Phi_k^0\right]\bigcup\left(j-1=1\right)\right\}$ \tabularnewline
\hline 
 $\mathcal{\overline P}_{f,h}^{10}$ & 0  & $\left(j\in \mathcal{S}^{10}\right) \bigcap \left[\left(j-1\right) \in \Phi_k^1\right]$   \tabularnewline
\hline 
 $\mathcal{\overline P}_d^{11}$ & $T^{11}$ & $\left(j\in \mathcal{S}^{11}\right) \bigcap \left\{\left[\left(j-1\right) \in \Phi_k^0\right]\bigcup\left(j-1=1\right)\right\}$ \tabularnewline
\hline 
 $\mathcal{\overline P}_{d,h}^{11}$ & 0 &  $\left(j\in \mathcal{S}^{11}\right) \bigcap \left[\left(j-1\right) \in \Phi_k^1\right]$ \tabularnewline
\hline 
 $\mathcal{\overline P}_d^{01}$ & $T^{01}$ & $\left(j\in \mathcal{S}^{01}\right) \bigcap \left\{\left[\left(j-1\right) \in \Phi_k^0\right]\bigcup\left(j-1=1\right)\right\}$ \tabularnewline
\hline 
 $\mathcal{\overline P}_{d,h}^{01}$ & 0 & $\left(j\in \mathcal{S}^{01}\right) \bigcap \left[\left(j-1\right) \in \Phi_k^1\right]$ \tabularnewline
\hline
\end{tabular}
\end{table} 

In summary, we have derived $\mathcal{T}^{i_0} \!\!\left\{\mathcal{S}_i\right\}$, which can be substituted into (\ref{EQN_T_T}) to obtain $\mathcal{T}^{i_0}$. Finally, 
applying the result of $\mathcal{T}^{i_0}$ to (\ref{EQN_T_put}), we can calculate the secondary network throughput $\mathcal{NT}$.

\section{Configuration of MAC Protocol for Throughput Maximization}

\subsection{Problem Formulation}
\label{TputOpt}

We are interested in determining optimal configuration of the proposed MAC protocol to achieve the maximum throughput while
satisfactorily protecting the PU. Specifically, let $\mathcal{NT}(T, W, P_s)$ denote the normalized secondary throughput,
which is the function of fragment time $T$, contention window $W$, and SU's transmit power $P_s$. 
Suppose that the PU requires that the average detection probability achieved at fragment $i$ be at least $\overline{\mathcal{P}}_{d,i}$. 
Then, the throughput maximization problem can be stated as follows:


\begin{equation}
\label{eq3a}
\begin{array}{l}
 {\mathop {\max }\limits_{T, W, P_s}} \quad {\mathcal{NT}} \left(T, W, P_s\right)  \\ 
 \mbox{s.t.}\,\,\,\, \hat{\mathcal{P}}_{d,i}\left(\varepsilon^i,T\right) \geq \mathcal{\overline P}_{d,i}, \quad i=1, 2,\cdots, \overline K \\
 \quad \quad 0 < T  \le T_{\sf eva},  \quad 0 < W \leq W_{\sf max}, \\
 \quad \quad 0 < P_s \leq P_{\sf max}, \\
 \end{array}\!\!
\end{equation}
where $W_{\sf max}$ is the maximum contention window, $P_{\sf max}$ is the maximum power for SUs and the fragment time $T$ is upper bounded by $T_{\sf eva}$.
In fact, the first constraint on $\hat{\mathcal{P}}_{d,i}\left(\varepsilon^i,T\right)$ implies that the spectrum sensing should be sufficiently
reliable to protect the PU where the fragment time (also sensing time) $T$ must be sufficiently large. Moreover, the optimal contention window $W$ should 
be set to balance between reducing collisions among SUs and limiting protocol overhead. Finally, the SU's transmit power $P_s$ must be appropriately
set to achieve good tradeoff between the network throughput and self-interference mitigation in spectrum sensing. 

\subsection{Configuration Algorithm for MAC Protocol}

\begin{algorithm}[h]
\scriptsize
\caption{\textsc{MAC Configuration Algorithm}}
\label{OPT_Throughput}
\begin{algorithmic}[1]

\FOR {each value of $W \in [1,W_{\sf max}]$}

\FOR {each searched value of $T \in \left(0,T_{\sf eva}\right]$}

\STATE Find optimal $P_s^*$ as $P_s^* = \mathop {\argmax} \limits_{0 \leq P_s \leq P_{\sf max}} \mathcal{NT} \left(T, W, P_s\right)$.

\ENDFOR

\STATE The best $\left(T^*, P_s^*\right)$ for each $W$ is $\left(T^*, P_s^*\right) = \mathop {\argmax} \limits_{T, P_s^*} \mathcal{NT} \left(T, W, P_s^*\right)$.
\ENDFOR

\STATE The final solution $\left(W^*, T^*, P_s^*\right) $ is determined as $\left(W^*, T^*, P_s^*\right) = \mathop {\argmax} \limits_{W, T^*, P_s^*} \mathcal{NT} \left(W, T^*, P_s^*\right)$.

\end{algorithmic}
\end{algorithm}

We assume the shifted exponential distribution for ${\tau}_{\sf ac}$ and ${\tau}_{\sf id}$ where ${\bar \tau}_{\sf ac}$ and ${\bar \tau}_{\sf id}$ are their corresponding 
average values of the exponential distribution.
Specifically, let $f_{\tau_{\sf x}}\left(t\right)$ denote the pdf of $\tau_{\sf x}$ (${\sf x}$ represents ${\sf ac}$ or ${\sf id}$ as we calculate the pdf of $\tau_{\sf ac}$ or $\tau_{\sf id}$, respectively) then
\beqn
f_{\tau_{\sf x}}\left(t\right) = \left\{{\begin{array}{*{20}{c}}
 \frac{1}{{\bar \tau}_{\sf x}} \exp(-\frac{t-T_{\sf min}^{\sf x}}{{\bar \tau}_{\sf x}}) & {\sf if} \, t \geq T_{\sf min}^{\sf x}\\ 
 0 & {\sf if} \, t <T_{\sf min}^{\sf x} 
\end{array}}\right.
\eeqn

For a given $T$, we would set the sensing detection threshold $\varepsilon$ and SU's transmit power $P_s$ so that the constraint on the average detection probability 
is met with equality, i.e., $\mathcal{\hat P}_d \left(\varepsilon,T\right) = \mathcal{\overline P}_d$ as in \cite{Liang08, Tan11}. 
In addition, the first constraint in (\ref{eq3a}) now turns to the two constraints for the average probabilities of detection under FD and HD spectrum sensing.
First, the average probability of detection under FD sensing can be expressed as
\beqn
\mathcal{\hat P}_d = \frac{\mathcal{P}_d^{11} \mathcal{P}\left(\mathcal{H}_{11}\right)+\mathcal{\overline P}_d^{01} \mathcal{P}\left(\mathcal{H}_{01}\right)}{\mathcal{P}\left(\mathcal{H}_{11}\right)+\mathcal{P}\left(\mathcal{H}_{01}\right)} 
\eeqn
where $\mathcal{P}_d^{11}$ is the probability of detection for $\mathcal{H}_{11}$; $\mathcal{\overline P}_d^{01}$ is the average probability of detection for $\mathcal{H}_{01}$, which
is given as
\beqn
\mathcal{\overline P}_d^{01} \!\! = \! \!\! \int_{T_{\sf min}^{\sf id}}^{T+T_{\sf min}^{\sf id}}  \mathcal{P}_d^{01}(t) \!\! f_{\tau_{\sf id}}\!\!\left(t\left|T_{\sf min}^{\sf id} \leq t \leq T+T_{\sf min}^{\sf id}\right.\right) dt 
\eeqn
where $f_{\tau_{\sf id}}\left(t\left|\mathcal{A}\right.\right)$ is the pdf of $\tau_{\sf id}$ conditioned on $\mathcal{A} = T_{\sf min}^{\sf id} \leq t \leq T+T_{\sf min}^{\sf id}$, which is given as
\beqn
f_{\tau_{\sf id}}\left(t\left|\mathcal{A}\right.\right) = \frac{f_{\tau_{\sf x}}\left(t\right)}{\Pr\left\{\mathcal{A}\right\}} = \frac{\frac{1}{{\bar \tau}_{\sf id}} \exp(-\frac{t}{{\bar \tau}_{\sf id}})}{1-\exp(-\frac{T}{{\bar \tau}_{\sf id}})}.
\eeqn
Note that $\mathcal{P}_d^{11}$ and $\mathcal{P}_d^{01}(t)$ are derived in Appendix \ref{CAL_P_F_P_D}. Moreover,
$\mathcal{P} \left(\mathcal{H}_{01}\right)$ and $\mathcal{P} \left(\mathcal{H}_{11}\right)$ are the probabilities of events $\mathcal{H}_{01}$ and $\mathcal{H}_{11}$, which are given as
\beqn
\mathcal{P} \!\! \left(\!\mathcal{H}_{01}\!\right) \!\!= \!\!\mathcal{P} \!\!\left(\!\mathcal{H}_0\!\right)\!\! \Pr \!\!\left(T_{\sf min}^{\sf id} \!\!\leq \!\!\tau_{\sf id} \!\!\leq \!\!T\!+\!T_{\sf min}^{\sf id}\!\right) \!\!=\!\! \mathcal{P}\!\! \left(\!\mathcal{H}_0\!\right)\!\! \left[\!1\!-\!\exp(\!\frac{-T}{{\bar \tau}_{\sf id}})\!\right] \label{PROB_H01} \\
\mathcal{P} \left(\!\mathcal{H}_{11}\!\right) \!=\! \mathcal{P} \!\!\left(\!\mathcal{H}_1\!\right)\! \Pr\!\! \left(\tau_{\sf ac} \!\!\geq\!\! T\!+\!T_{\sf min}^{\sf ac}\right)\! =\!  \mathcal{P} \left(\!\mathcal{H}_1\!\right)  \exp(-\frac{-T}{{\bar \tau}_{\sf ac}})
\eeqn
The average probability of detection for HD sensing, $\mathcal{\hat P}_{d,h}$ can be derived similarly.

We propose an algorithm to determine $\left(T, W, P_s\right)$ summarized in Alg.~\ref{OPT_Throughput}.
Note that there is a finite number of values for $W \in [1,W_{\sf max}]$; therefore, we can perform exhaustive search to determine its best value.
Moreover, we can use the bisection scheme to determine the optimal value of $T$. Furthermore, the optimal value of  $P_s$ can be determined by a numerical method
for given $T$ and $W$ in step 3. Then, we search over all possible choices of $T$ and $W$ to determine the optimal configuration of the parameters (in steps 5 and 7).

\subsection{Half-Duplex MAC Protocol with Periodic Sensing}
\label{HDMACDESIGN}

To demonstrate the potential performance gain of the proposed FD MAC protocol, we also consider an HD MAC protocol.
In this HD MAC protocol, we perform the same backoff for channel resolution but we employ
periodic HD sensing in each data fragment where the sensing duration is $T_S$ and data transmission duration is $T-T_S$ . 
If the sensing outcome in the sensing stage indicates an available channel then the SU transmit data in the second stage; 
otherwise, it will keep silent for the remaining time of fragment and wait for the next fragment.
Due to the space constraint, throughput analysis for this HD MAC protocol is given in the online
technical report \cite{report}.

\section{Numerical Results}
\label{Results}

To obtain numerical results, we take key parameters for the MAC protocol from Table II in \cite{Bian00}. All other parameters are chosen as follows unless stated otherwise:
mini-slot is $\sigma = 20 {\mu} s$; sampling frequency is $f_s = 6$MHz; bandwidth of PU's QPSK signal is $6$MHz; $\mathcal{\overline P}_d = 0.8$; $T_{\sf eva} = 40$ms; $T_{\sf min}^{\sf ac} = 40$ms; $T_{\sf min}^{\sf id} = 45$ms; the SNR of PU 
signals at SUs $\gamma_P = \frac{P_p}{N_0} = -20$dB; the self-interference parameters $\zeta = 0.4$ and varying $\xi$. Without loss of generality, the noise power is
 normalized to one; hence, the SU transmit power, $P_s$ becomes $P_s = SNR_s$; and $P_{\sf max} = 25$dB.

\begin{figure}[!t]
\centering
\includegraphics[width=50mm]{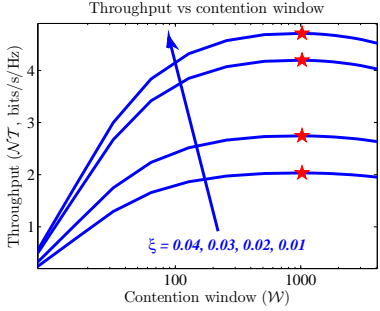}
\caption{Normalized throughput versus contention window $W$  for $T = 18 ms$, ${\bar \tau}_{\sf id} = 1000 ms$, ${\bar \tau}_{\sf ac} = 100 ms$, $\overline{K} = 4$ and varying $\xi$.}
\label{T_VS_CW_xi_1axis}
\end{figure}

\begin{figure}[!t]
\centering
\includegraphics[width=50mm]{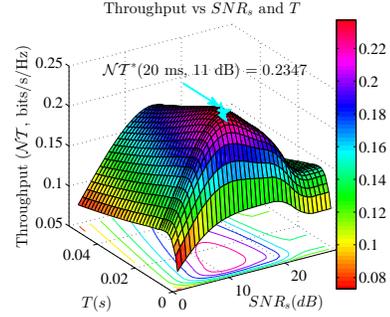}
\caption{Normalized throughput versus SU transmit power $P_s$ and length of fragment $T$ for $W = 1024$, ${\bar \tau}_{\sf id} = 200 ms$, $n_0 = 40$, ${\bar \tau}_{\sf ac} = 100 ms$, $\overline{K} = 4$, $\xi =0.95$ and $\zeta = 0.6$.}
\label{T_vs_Ps_tau}
\end{figure}

\begin{figure}[!t]
\centering
\includegraphics[width=50mm]{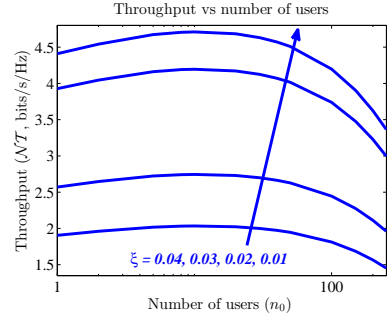}
\caption{Normalized throughput versus number of SUs $n_0$ for $T = 18 ms$, $W = 1024$, ${\bar \tau}_{\sf id} = 1000 ms$, ${\bar \tau}_{\sf ac} = 100 ms$, $\overline{K} = 4$ and varying $\xi$.}
\label{T_VS_n0_xi_1axis1}
\end{figure}


\begin{figure}[!t]
\centering
\includegraphics[width=50mm]{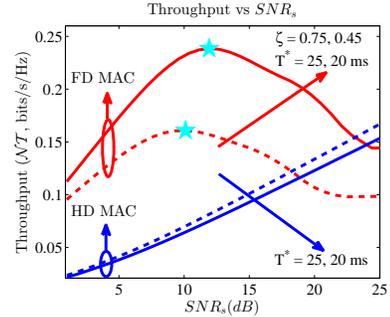}
\caption{Normalized throughput versus $P_s$ for $\left({\bar \tau}_{\sf id}, {\bar \tau}_{\sf ac}\right) = \left(200, 100\right)$ ms, $n_0 = 40$, $\overline{K} = 4$, $\xi =0.9$ and $\zeta = \left\{0.45, 0.75\right\}$.}
\label{T_vs_Ps_FD_HD}
\end{figure}

We first consider the effect of self-interference on the throughput performance where $\zeta = 0.4$ and
$\xi$ is varied in $\xi = \left\{0.01, 0.02, 0.03, 0.04\right\}$.
Fig.~\ref{T_VS_CW_xi_1axis} illustrates the variations of the throughput versus contention window $W$.
It can be observed that when $\xi$ decreases (i.e., the self-interference is smaller), the achieved throughput increases.
This is because the SU can transmit with higher power while still maintaining the sensing constraint, which leads to throughput improvement. 
The optimal $P_s$ corresponding to these values of $\xi$ are $P_s = SNR_s = \left\{25.00, 18.19, 13.56, 10.78\right\}$dB and 
the optimal contention window  is indicated by a star.

Fig.~\ref{T_vs_Ps_tau} illustrates the throughput performance versus SU transmit power $P_s$ and length of fragment $T$ where $\xi = 0.95$,
 $\zeta = 0.45$, $T_{\sf eva} = T_{\sf min}^{\sf ac} = T_{\sf min}^{\sf id} = 40$ms, $P_{\sf max} = 30$dB.
It can be observed that there exists an optimal configuration of SU transmit power $P_s^* = 11 dB$ and  fragment time $T^* = 20 ms$ to achieve the
 maximum throughput $\mathcal{NT}\left(T^*, P_s^*\right) = 0.2347$, which is indicated by a star symbol. This demonstrates
the significance of power allocation to mitigate the self-interference and the optimization of fragment time $T$ to effectively
exploit the spectrum opportunity. 
Fig.~\ref{T_VS_n0_xi_1axis1} illustrates the throughput performance versus number of SUs, $n_0$ where  $\zeta = 0.4$
and $\xi$ is varied in the set $\xi = \left\{0.01, 0.02, 0.03, 0.04\right\}$.
Again, when $\xi$ decreases (i.e., the self-interference is smaller), the achieved throughput increases. 
In this figure, the $SNR_s$ corresponding to the considered values of $\xi$ are $P_s = SNR_s = \left\{25.00, 18.19, 13.56, 10.78\right\} dB$.

Finally, we compare the throughput of our proposed FD MAC protocol and the HD MAC protocol with periodic sensing in Fig.~\ref{T_vs_Ps_FD_HD}.
For fair comparison, we first obtain the optimal configuration of FD MAC protocol, i.e., $\left(T^*, W^*, P_s^*\right)$ ($\left(20 ms, 1024, 12 dB\right)$ for 
$\zeta = 0.45$ and $\left(25 ms, 1024, 10 dB\right)$ for $\zeta = 0.75$), then we use $\left(T^*, W^*\right)$ for the HD MAC protocol.
Moreover, we optimize sensing time, $T_S$ to maximize the achieved throughput for the HD MAC protocol.
We can see that when the self-interference is higher (i.e., $\zeta$ increases), the FD MAC protocol requires higher fragment length, $T$ but lower 
SU transmit power, $P_s$.
For both studied cases of $\zeta$, our proposed FD MAC protocol with power allocation
 outperforms the HD MAC protocol at the corresponding optimal power levels
required by the FD MAC protocol. Moreover, we can observe that our proposed FD MAC protocol can achieve the maximum throughput at the transmit 
power level less than $P_{\sf max}$  while the HD MAC protocol achieves the maximum throughput at $P_{\sf max}$ since it does not suffer from the self-interference.

\section{Conclusion}
\label{conclusion} 
In this paper, we have proposed the FD MAC protocol for FDCRNs that explicitly takes into account the self-interference.
Specifically, we have derived the normalized throughput of the proposed MAC protocols and determined their optimal configuration for throughput 
maximization. Finally, we have presented numerical results to demonstrate the desirable performance of the proposed design.

\appendices

\section{Probabilities of a False Alarm and Detection}
\label{CAL_P_F_P_D}

Assume that the transmitted signals from the PU and SU transmitter are circularly symmetric complex Gaussian (CSCG) signals while the noise at the secondary links is independent
and identically distributed CSCG $\mathcal{CN}\left( {0,{N_0}} \right)$ \cite{Liang08}. 
Under FD sensing, the probability of a false alarm for event $\mathcal{H}_{00}$ can be derived using the similar method as the one in \cite{Liang08},
which is given as
\beqn
\mathcal{P}_f^{00} = \mathcal{Q} \left[\left(\frac{\epsilon}{N_0+I}-1\right)\sqrt{f_sT}\right]
\eeqn
where $\mathcal{Q} \left(x\right) = \int_x^{+\infty} \exp \left(-t^2/2\right) dt$; $f_s$, $N_0$, $\epsilon$, $I$ are the sampling frequency, the noise power, the detection
threshold and the self-interference, respectively. The probability of detection for event $\mathcal{H}_{11}$ is
\beqn
\mathcal{P}_d^{11} = \mathcal{Q} \left[\left(\frac{\epsilon}{N_0+I}-\gamma_{PS}-1\right)\frac{\sqrt{f_sT}}{\gamma_{PS}+1}\right]
\eeqn
where $\gamma_{PS} = \frac{P_p}{N_0+I}$ is the signal-to-interference-plus-noise ratio (SINR) of the PU's signal at the SU. 

Similarly, we can express $\mathcal{P}_f^{10}$ and $\mathcal{P}_d^{01}$ as follows:
\beqn
\mathcal{P}_f^{10} =  \mathcal{Q} \!\! \left(\!\frac{\left(\frac{\epsilon}{N_0+I}- \frac{t}{T}\gamma_{PS}-1\right)\sqrt{f_sT}}{\sqrt{\frac{t}{T}\left(\gamma_{PS}+1\right)^2+1-\frac{t}{T}}}\!\!\right) \!\! \\
\mathcal{P}_d^{01} \!\! =  \mathcal{Q} \left( \!\!\frac{\left(\!\! \frac{\epsilon}{N_0+I}- \frac{T-t}{T}\gamma_{PS}-1\right)  \!\sqrt{f_sT}}{\sqrt{\frac{T-t}{T}\left(\gamma_{PS}+1\right)^2+\frac{t}{T}}} \!\! \right) 
\eeqn
where $t$ is the time instant when the PU changes its state.
For HD sensing, the expressions for the probabilities of detection and a false alarm for the corresponding four events are similar to the ones for FD sensing 
except that the self-interference-plus-noise power $N_0+I$ becomes noise power $N_0$ only; hence, $\gamma_{PS}$ becomes $\gamma_{PS}^h = \frac{P_p}{N_0}$. 


\section{Fragment Throughput}
\label{TPUT_FRAGMEN}

For $\mathcal{H}_{00}$, the average throughput $T^{00}$ is
\beqn
T^{00} = T \log_2 \left(1+\gamma_{S1}\right)
\eeqn
where $\gamma_{S1} = \frac{P_s}{N_0+I}$ is the SINR of received signal at the SU receiver when the PU is idle.
Similarly, we can write $T^{10}$, $T^{01}$ and $T^{11}$ as follows:
\beqn
T^{10} &=& \left[t \log_2 \!\!\left(1+\gamma_{S2}\right) \!+\! (T-t) \log_2 \!\!\left(1+\gamma_{S1}\right)\right] \\
T^{01} &=& \left[t \log_2 \!\!\left(1+\gamma_{S1}\right) \!+\! (T-t) \log_2 \!\!\left(1+\gamma_{S2}\right)\right]  \\
T^{11} &=& T \log_2 \left(1+\gamma_{S2}\right)
\eeqn
where $\gamma_{S2} = \frac{P_s}{N_0+I+P_p}$ is the SINR of the received signal at the SU receiver when the PU is active, and $t$ is the time instant at which the PU changes its activity state.

\bibliographystyle{IEEEtran}


\begin{thebibliography}{19}


\bibitem{Yu09}
T.~Yucek and H.~Arslan, ``A survey of spectrum sensing algorithms for cognitive radio applications,'' {\em IEE Commun. Surveys and Tutorials,} vol. 11, no. 1, pp. 116--130, 2009.
 
\bibitem{Cor09}
C. Cormiob and K. R. Chowdhurya, ``A survey on MAC protocols for cognitive radio networks,'' {\em Elsevier Ad Hoc Networks,} 
vol. 7, no. 7, pp. 1315-1329, Sept. 2009. 

\bibitem{Liang08}
 Y.~ C.~ Liang, Y.~ H.~ Zeng, E.~ C.~ Y.~ Peh, and A.~ T.~ Hoang, ``Sensing-throughput tradeoff for cognitive radio networks,'' {\em IEEE Trans. Wireless Commun.}, vol. 7, no. 4, pp. 1326--1337, April 2008.

\bibitem{Tan11}
L.~ T.~ Tan and L.~ B.~ Le, ``Distributed MAC protocol for cognitive radio networks: Design, analysis, and optimization,'' {\em IEEE Trans. Veh. Technol.}, vol. 60, no. 8, pp. 3990--4003, Oct. 2011.

\bibitem{Tan12}
L.~ T.~ Tan and L.~ B.~ Le, ``Channel assignment with access contention resolution for cognitive radio networks,'' {\em IEEE Trans. Veh. Technol.}, vol. 61, no. 6, pp. 2808--2823, April 2012.

\bibitem{Konda08}
Y.R.~Kondareddy, and P.~Agrawal, ``Synchronized MAC protocol for multi-hop cognitive radio networks,'' in {\em Proc. IEEE ICC'}2008.


\bibitem{Afifi14}
W.~ Afifi and M.~ Krunz, ``Incorporating self-interference suppression for full-duplex operation in opportunistic spectrum access systems,'' 
 \emph{IEEE Trans. Wireless Commun.}, vol. 14, no. 4, April 2015.


\bibitem{Duarte12}
M. Duarte, C. Dick, and A. Sabharwal, ``Experiment-driven characterization of full--duplex wireless systems,'' \emph{IEEE Trans. Wireless Commun.}, vol. 11, no. 12, pp. 4296--4307, Dec. 2012.

\bibitem{Bian00}
G. Bianchi, ``Performance analysis of the ieee 802.11 distributed coordination function,'' {\em IEEE J. Sel. Areas Commun.}, vol. 18, no. 3, pp. 535-547, Mar. 2000.

\bibitem{report}
L. T. Tan and L. B. Le, ``Distributed MAC protocol design for full--duplex cognitive radio networks,'' technical report. 
Online: http://www.necphy-lab.com/pub/TanReport.pdf






%


%
%
%
%
%

%




%
%
%

  
%
%

    %
%



 


 





 



\end{thebibliography}

%
%
%
%
%
%
%


\end{document}